\documentclass[pra,superscriptaddress,twocolumn]{revtex4}
\usepackage{amssymb,amsthm,amsmath,color,graphics}

\newcommand{\ket}[1]{\ensuremath{\left|#1\right>}}

\newcommand{\ketbra}[2]{| #1 \rangle\langle #2 |}

\newcommand{\nchoosek}[2]{\left(\! \begin{array}{c}#1\\#2\end{array}\! \right)}
\newcommand{\floor}[1]{\lfloor #1 \rfloor}

\newcommand{\be}{\begin{eqnarray}}
\newcommand{\ee}{\end{eqnarray}}

\newcommand{\tr}{\text{tr}}

\newcommand{\ip}[2]{\tr(#1 #2)}
\def\opone{\leavevmode\hbox{\small1\kern-3.8pt\normalsize1}}

\begin{document}
\title{An improved algorithm for quantum separability and entanglement detection}

\author{L. M. Ioannou}
\email{lmi22@cam.ac.uk}
\affiliation{Centre for Quantum
Computation, Department of Applied Mathematics
  and Theoretical Physics, University of Cambridge, Wilberforce Road, Cambridge
  CB3 0WA, UK}

\author{B. C. Travaglione}
\affiliation{Centre for Quantum
Computation, Department of Applied Mathematics
  and Theoretical Physics, University of Cambridge, Wilberforce Road, Cambridge
  CB3 0WA, UK}
\affiliation{Computer Laboratory, University of Cambridge, JJ Thomson Ave, Cambridge CB3 0FD, UK}

\author{D. Cheung}
\affiliation{Combinatorics and Optimization, University of
Waterloo, 200 University Avenue West, Waterloo, N2L 3G1, Canada}

\author{A. K. Ekert}
\affiliation{Centre for Quantum Computation, Department of Applied
Mathematics
  and Theoretical Physics, University of Cambridge, Wilberforce Road, Cambridge
  CB3 0WA, UK}

\begin{abstract}
Determining whether a quantum state is separable or entangled is a
problem of fundamental importance in quantum information science.
It has recently been shown that this problem is NP-hard.  There is
a highly inefficient `basic algorithm' for solving the quantum
separability problem which follows from the definition of a
separable state.  By exploiting specific properties of the set of
separable states, we introduce a new classical algorithm that
solves the problem significantly faster than the `basic
algorithm', allowing a feasible separability test where none
previously existed e.g. in 3-by-3-dimensional systems.  Our
algorithm also provides a novel tool in the experimental detection
of entanglement.
\end{abstract}

\maketitle


Entangled quantum states are interesting both from theoretical and
practical points of view.  Theoretically, entanglement is
connected to the confounding issue of nonlocality.  Practically,
entangled states are useful in quantum cryptography and other
quantum information processing tasks (see \cite{NC00} and
references therein). A mixed quantum state is defined as
\emph{separable} if and only if it can be written as a convex
combination of pure separable states (and defined as
\emph{entangled}, otherwise). Solving the quantum separability
problem simply means determining whether a given quantum state is
entangled or separable. The problem comes in two flavors -- one
theoretical, and the other experimental. In this paper, we
describe an algorithm for solving the quantum separability problem
in the theoretical setting.  We also describe the algorithm's
utility in the experimental setting.


We begin by introducing some notation and precisely defining the
quantum separability problem. In what follows, we are considering
a bipartite quantum system of dimension $M \times N$. Let
$\mathbb{H}_{M,N}$ denote the vector space of all Hermitian
operators acting on \mbox{$\mathbb{C}^M\otimes\mathbb{C}^N$}.
Noting that $\mathbb{H}_{M,N}$ is isomorphic to
$\mathbb{R}^{M^2N^2}$, it is endowed with the Euclidean
inner-product $\langle X, Y \rangle\equiv \tr(XY)$, which induces
the corresponding norm $||X||\equiv\sqrt{\tr(X^2)}$ and distance
measure $||X-Y||$. Let $\mathcal{D}_{M,N} \subset
\mathbb{H}_{M,N}$ denote the set of all density operators.  The
set of bipartite separable quantum states $\mathcal{S}_{M,N}
\subset \mathcal{D}_{M,N}$ is defined as the convex hull of the
separable pure states
$\{\ketbra{\alpha}{\alpha}\otimes\ketbra{\beta}{\beta}\}$, where
$\ket{\alpha}$ ($\ket{\beta}$) is a normalized vector in
$\mathbb{C}^M$ ($\mathbb{C}^N$).  An arbitrary density matrix in
$\mathcal{D}_{M,N}$ is parameterized by $n-1$ real variables,
where \be
 n \equiv M^2N^2,
\ee
and an arbitrary separable pure
state is parameterized by
\be
k \equiv 2(M+N)-4
\ee
real variables.
Of course, in defining the separability problem, we cannot allow
infinite precision, so we need to introduce a precision parameter
$\delta>0$.
We are now ready to define the \emph{(quantum) separability
problem} as
follows:\\
\emph{ {\bf Quantum Separability Problem.} Given a density matrix
$\rho \in \mathcal{D}_{M,N}$ and a precision $\delta$, assert
either
\begin{tabbing}
\hspace*{0.5cm}\= ENTANGLED \=:\ \=\kill
\> SEPARABLE: \>\hspace{1mm} there exists a separable state $\sigma$\\
\> \> \> such that $||\rho-\sigma|| < \delta$; or \\
\> ENTANGLED: \>\hspace{1mm}  there exists an entangled state $\tau$\\
\> \> \> such that $||\rho-\tau|| < \delta$.
\end{tabbing}
}


The separability problem has been shown to be NP-hard \cite{Gur03},
thus any devised test
for separability is likely to require a number of computing
resources that scales exponentially with $M$ and $N$. There exist
efficient ``one-sided'' tests for separability, where the output
of some polynomial-time computable function of the matrix for
$\rho$ can indicate that $\rho$ is certainly entangled
\cite{Per96,HH99,NK00,DPS01,qphDPS03} or certainly separable
\cite{BCJLPS99,ZHSL98,GB02}, but not both.


The experimental flavor of the separability problem can be defined
as follows:  Given many physical copies of a completely unknown
quantum state $\rho$, determine whether $\rho$ is separable.  One
may solve this problem by performing full state tomography in
order to construct the density matrix for $\rho$ to some precision
$\delta$, and then solve the theoretical separability problem.  If
there is some partial knowledge of $\rho$, then there are more
options, such as testing for a violation of a specific Bell
inequality \cite{Bel64,CHSH69} or invoking entanglement witnesses
\cite{GHBELMS02,qphBMNMDM03}. In the case where $MN\leq 6$, the
positive partial transpose (PPT) test \cite{Per96,HHH96} can be
implemented physically \cite{qphHE01,qphCar03}, though currently
this approach is not experimentally viable.


The `basic algorithm' that follows from the definition of a
separable state is simply a straightforward search for a convex
combination of separable pure states that gives the required
density matrix within precision $\delta$.
Since any separable density operator in $\mathcal{S}_{M,N}$ can be
written as a convex combination of $n$ separable pure states
\cite{H97}, a lower bound for the worst-case run time, $t_1$, of
this search is given by \be\label{BasicRunTime} t_1(n,\delta) &>&
\nchoosek{\Omega_{\delta}}{n}\nchoosek{\floor{1/\delta}}{n-1}
\times \text{poly}(n,\log(1/\delta)), \ee where $\Omega_{\delta}$
is the number of pure separable states to precision $\delta$. The
first binomial factor is a lower bound for the number of
combinations of $n$ distinct pure separable states to precision
$\delta$; the second is a lower bound for the number of
probability distributions over the $n$ states to precision
$\delta$. The algorithm that we present here has a worst-case run
time, $t_2$, with an upper bound of \be\label{OurRunTime}
t_2(n,\delta) &<& \Omega_\delta \times
\text{poly}(n,\log(1/\delta)). \ee To compare these run times it
is not necessary to compute the exact value of $\Omega_{\delta}$,
but note that for realistic values of $\delta$, $\Omega_\delta$
will be of order $O(2/\delta^k) \gg n$, so $t_2$ will be
significantly less than $t_1$. Later it will be explained that
equation (\ref{OurRunTime}) can be improved further in practice
with the use of global optimization routines, making the algorithm
of practical use in the case where $M$ and $N$ are small (and
$\delta$ is not too small). Even for $M=N=3$, there was previously
no known better algorithm for the separability problem than the
basic one described above. Note also that for $MN\leq 6$, where
the PPT test is necessary and sufficient, our algorithm still
offers its novel advantage in the experimental setting (as
explained later).


Before describing our algorithm for separability, we note the
following fact \cite{HHH96}: A state $\rho$ is entangled if and
only if there exists an entanglement witness \cite{Ter00} that
detects it. An \emph{entanglement witness} is any traceless
operator $A\in\mathbb{H}_{M,N}$ for which there exists a state
$\rho\in\mathcal{D}_{M,N}$ such that
\begin{eqnarray}
\tr(A\sigma)&<&\tr(A\rho)\quad \forall \sigma \in
\mathcal{S}_{M,N}.
\end{eqnarray}
This definition is slightly different from that used in the
literature, however it substantially simplifies the description of
the algorithm. Recalling that $\mathbb{H}_{M,N}$ is isomorphic to
$\mathbb{R}^{n},$ the above definition implies that for entangled
$\rho$ there exists a hyperplane which separates $\rho$ from the
set of all separable states $\mathcal{S}_{M,N}$. If one defines
the function
\begin{eqnarray}\label{bA}
b_A &\equiv& \max_{\sigma\in \mathcal{S}_{M,N}}\tr(A\sigma),
\end{eqnarray}
then the set $\lbrace X\in\mathbb{H}_{M,N}:\hspace{2mm}
\ip{A}{X}=b_A\rbrace$ is one such hyperplane.
We (non-uniquely) define $\sigma_A$ to be any element of
$\mathcal{S}_{M,N}$ such that $\tr(A\sigma_A)=b_A$. It suffices
only to consider entanglement witnesses $A$ such that
$\tr(A^2)=1$, that is, those which lie on the
$(n\!-\!2)$-dimensional surface of a $\mathbf{0}$-centered
unit-hypersphere in $\mathbb{H}_{M,N}$, where $\mathbf{0}$ is the
origin (null operator in $\mathbb{H}_{M,N}$).  For our purposes,
however, it will be useful to characterize \emph{all potential
entanglement witnesses} by the corresponding $(n-1)$-dimensional
\emph{unit-hyperball}, $\mathcal{W}$, defined as
\begin{eqnarray}
\mathcal{W}\equiv \{A\in\mathbb{H}_{M,N}: \tr(A)=0,\tr(A^2)\leq
1\}.
\end{eqnarray}
 We can now define the \emph{(entanglement)
witness problem},
a problem slightly harder than the separability problem: \\
\emph{ {\bf Entanglement Witness Problem.} Given a density matrix
$\rho \in \mathcal{D}_{M,N}$ and a precision $\delta$, either assert
\begin{tabbing}
\hspace*{0.4cm}\= ENTANGLED \=:\ \=\kill
\>   SEPARABLE: \>there exists a separable state $\sigma$\\
\> \> such that $||\rho-\sigma|| < \delta$; or return \\
\> $A\in\mathcal{W}$: \> an operator such that \\
\> \> $\tr(A\sigma) < tr(A\rho) +  \delta$ for all $\sigma \in
\mathcal{S}_{M,N}$.
\end{tabbing}
} \noindent The witness problem is thus to decide that $\rho$ is
almost separable, or to find an approximate entanglement witness
for $\rho$. Note that any algorithm solving the witness problem
also solves the separability problem. Our algorithm actually
solves the witness problem.


Our algorithm is an iterative one, which calls a computationally
expensive subroutine at each iteration. It is convenient to treat
this subroutine as a black box, or oracle, when describing the
algorithm's main structure.  Simply define the \emph{oracle},
$\mathcal{O}$, such that it takes an operator $A$, and returns
$\mathcal{O}(A)\equiv\sigma_A$.  Whichever way $\mathcal{O}(A)$ is
computed, it suffices that the maximization in (\ref{bA}) is done
over the \emph{pure} separable states (and that $\sigma_A$ is a
pure state).  Thus, the motivation behind reducing the
separability problem to the oracle $\mathcal{O}$ is quite simple:
we are exploiting the fact that the separable pure states, which
are the extreme points of $\mathcal{S}_{M,N}$, are parameterized
by $k$ variables rather than $n$.  Thus, from a practical point of
view, the complexity of computing $\mathcal{O}(A)$ scales much
better than that of either a brute-force search through all
entanglement witnesses, or the `basic algorithm'.


Before delving into the details, we give a high-level description
of our algorithm.  The algorithm maintains a set
$\mathcal{K}\subseteq\mathcal{W}$ of operators which are potential
entanglement witnesses for $\rho$. If (and only if) $\rho$ is
entangled, there exists a closed, convex subset of $\mathcal{W}$,
which we call $\mathcal{W}_{\rho}$, consisting of \emph{all
entanglement witnesses that detect $\rho$}.
Throughout the algorithm, we have $\mathcal{W}_\rho\subset
\mathcal{K}$. Initially, $\mathcal{K}$ is set equal to
$\mathcal{W}$. In each iteration, the algorithm selects a
\emph{test-witness}, $A\in\mathcal{K}$, and computes
$\sigma_A=\mathcal{O}(A)$.  If $\tr(A\sigma_A) <
\tr(A\rho)+\delta$, then the algorithm returns $A$; else,
$\mathcal{K}$ is reduced and the next iteration begins.  The
algorithm keeps reducing the set $\mathcal{K}$ until it either
finds an (approximate) element of $\mathcal{W}_{\rho}$, or it
decides that $\mathcal{W}_{\rho}$ is empty, and therefore $\rho$
is separable. To decide $\mathcal{W}_\rho$ is empty means that
$\mathcal{K}$ is too small to contain $\mathcal{W}_\rho$.  This
requires having a lower bound on the size of $\mathcal{W}_\rho$.
By exploiting the role of $\delta$ in the problem definitions,
such a lower bound can be derived in terms of $\delta$ and $n$.
In what follows, we will ignore
the role of $\delta$, as it obfuscates the main idea of the
algorithm.


We now describe how to reduce the set $\mathcal{K}$, that is, to
discard elements of $\mathcal{K}$ that are not elements of
$\mathcal{W}_\rho$. Suppose $A$ is not in $\mathcal{W}_{\rho}$,
but is sufficiently close to $\mathcal{W}_{\rho}$. Then, $A$,
$\rho$, and $\sigma_A$ can be used to define a half-space
$\{X\in\mathbb{H}_{M,N}:\ip{K}{X}\geq 0\}$ that contains
$\mathcal{W}_\rho$.
Specifically, we have the following: Let $W$ be any operator in
$\mathcal{W}_{\rho}$ and suppose $A \notin \mathcal{W}_{\rho}$. If
$\ip{W}{A} \geq 0$, then choosing
\begin{eqnarray}\label{Lemma}
K &\equiv& (\rho-\sigma_A) -
\frac{\ip{A}{(\rho-\sigma_A)}}{\tr(A^2)}A
\end{eqnarray}
(and then normalizing $K$) gives $\ip{K}{A}=0$ by construction,
and it is easy to verify that $\ip{K}{W}>0$. The idea is that, at
each iteration, the test-witness $A$ is chosen so that it is
(approximately) in the center of the current $\mathcal{K}$
(relative to the Euclidean geometry). If the oracle $\mathcal{O}$
returns $\sigma_A$ such that $\tr(A\rho)<\tr(A\sigma_A)$, then, as
long as $\ip{W}{A}\geq 0$ for all $W\in\mathcal{W}_\rho$, equation
(\ref{Lemma}) gives a \emph{cutting plane}
$\{X\in\mathbb{H}_{M,N}:\ip{K}{X}=0\}$ that slices through $A$ and
$\mathbf{0}$.  This allows us to discard the half of $\mathcal{K}$
consisting of operators $X$ such that $\ip{K}{X}\leq 0$. Because
$\mathcal{K}$ is being approximately halved at each step, the
algorithm quickly either finds an entanglement witness for $\rho$
or concludes that $\rho$ is separable.


Our problem of determining whether the convex set
$\mathcal{W}_\rho$ is empty using cutting planes is well studied
in the field of convex optimization.  However, because of our
special requirement that $\ip{W}{A}\geq 0$ for all
$W\in\mathcal{W}_\rho$, none of the existing algorithms can be
applied directly.  Fortunately, though, the
analytic-central-section algorithm due to Atkinson and Vaidya
\cite{AV95} can be adapted for our purpose, giving an algorithm
with the desired complexity.


We now describe the algorithm.  Let $I_{MN}$ be the maximally
mixed state, which is properly contained in $S_{M,N}$
\cite{BCJLPS99,GB02}. It is easy to verify that
$\mathcal{W}_{\rho}$ must be contained in the half-space $\lbrace
X:\ip{(\rho-I_{MN})}{X}\geq 0\rbrace$. Let $K_1 \equiv
(\rho-I_{MN})/||\rho-I_{MN}||$. Thus, straight away, $\mathcal{K}$
is reduced to the half-ball $\mathcal{W}\cap\lbrace
X:\ip{K_1}{X}\geq 0\rbrace$. The first test-witness to give to the
oracle is $A=\rho-I_{MN}$ (which is along the center-line of the
half-ball). If the oracle confirms that $A$ detects $\rho$, then
we are done. Otherwise, we use equation (\ref{Lemma}) to generate
a cutting plane. By way of mathematical induction, assume that, at
some later stage in the algorithm, $\mathcal{K}$ has been reduced
to
\begin{eqnarray}
\mathcal{K} &=& \mathcal{W}\bigcap \cap_{i=1}^{h}\lbrace
X:\ip{K_i}{X}\geq 0 \rbrace,
\end{eqnarray}
by the generation of $h$ cutting planes $\lbrace X: \ip{K_i}{X}=0
\rbrace$, as described above. Recall that we want to choose a
test-witness that is approximately in the center of $\mathcal{K}$.
An easily computable candidate is the \emph{analytic center}, $C$,
of $\mathcal{K}$ \cite{NN94}, which is defined as the unique
minimizer of the real convex function
\begin{eqnarray}
F(X) &\equiv& -\sum_{i=1}^{h}\log(\ip{K_i}{X}) -
\log(1-||X||^2),\hspace{4mm}
\end{eqnarray}
defined for $X\in\mathcal{K}$.  The relation $\nabla F(C)=0$ gives
$C=\frac{1-||C||^2}{2}\sum_{i=1}^{h}\frac{K_i}{\ip{K_i}{C}}$,
which, by the inductive hypothesis, implies that $\ip{W}{C}\geq 0$
for all $W\in \mathcal{W}_p$. Thus, $A=C$ is a suitable
test-witness to give to the oracle and to use in equation
(\ref{Lemma}). Full details of a robust algorithm are too numerous
to include here but can be derived with the help of
\cite{AV95,Ren01,NN94}. The important point is that the
separability of a given density matrix can be decided with only
$n\times\text{polylog}\left(n,1/\delta\right)$ calls
to the oracle.


Now consider the complexity of computing $\mathcal{O}(A)$, which
so far has been black-boxed.  The most na\"{i}ve way to carry out
this computation is to one-by-one calculate $\tr(A\sigma)$ for
each of the $\Omega_{\delta}$ pure separable states $\sigma$ (to
precision $\delta$) and return the $\sigma$ that produced the
largest value of $\tr(A\sigma)$. Even with this na\"{i}ve way of
computing $\mathcal{O}(A)$, the total run time of our algorithm is
significantly shorter than that of the `basic algorithm' for
quantum separability (compare (\ref{BasicRunTime}) and
(\ref{OurRunTime})). However, for any given orthogonal Hermitian
basis of $\mathbb{H}_{M,N}$, the closed, general form of the
function $\tr(A\sigma)$ can be written down in terms of the $k$
real parameters of the separable pure states. Armed with the
closed form of the function to be maximized, various well-studied
global maximization techniques are at one's disposal,
for example, Lipschitz optimization \cite{HP95} or interval analysis
\cite{HW04}. Call the function to be maximized $f$ and denote its
global maximum by $f^*$.  As the global optimization algorithm
proceeds, it gives progressively better lower and upper bounds on
$f^*$. Call these bounds $\underline{f}$ and $\overline{f}$,
respectively.  A key advantage of our algorithm is that, during
any computation of $\mathcal{O}(A)$, the search for $f^*$ may be
halted early when either (\emph{i}) $\tr(A\rho) \leq
\underline{f}$, in which case equation (\ref{Lemma}) can be
invoked to generate a new cutting plane, or (\emph{ii})
$\overline{f}< \tr(A\rho)$, in which case the algorithm has found
an entanglement witness for $\rho$. Thus, the algorithm's run time
may be significantly shorter than the worst-case analysis
predicts.


Finally, we discuss how the algorithm may be used when only
partial information about the state $\rho$ is available. This is
of particular use in an experimental setting.  Let $\mathcal{B}$
be an orthonormal, Hermitian basis for $\mathbb{H}_{M,N}$. The
state $\rho$ can be written $\rho = \sum_{i=1}^n \rho_{X_i}X_i$,
where $\rho_{X_i}\in\mathbb{R}$. Each coefficient $\rho_{X_i}$ is
simply the \emph{expected value of $X_i$}, which equals
$\text{\tr}(X_i\rho)$. The expected values of all elements of
$\mathcal{B}$ constitute complete information about $\rho$.
Suppose we have only measured $j < n$ expected values. The
algorithm can be applied in this reduced, $j$-dimensional space.
If the algorithm finds a hyperplane separating $\rho$ from
$\mathcal{S}_{M,N}$, then $\rho$ is entangled; otherwise $\rho$
may be entangled or separable, as the $j$ expected values are
consistent with a separable state. As expected values are being
gathered through experimental observation, they may be input to
the algorithm. If the basis $\mathcal{B}$ is separable, then the
entire procedure can be done when the subsystems are spatially
separated with local operations and classical communication.  The
idea of searching for an entanglement witness in the span of
operators whose expected values are known was discovered
independently and applied, in a special case, to quantum
cryptographic protocols in \cite{qphCLL03}.


We have given a classical algorithm for the quantum separability
problem which takes as input the density matrix for a quantum
state $\rho$ and either decides that $\rho$ is separable, or
returns an entanglement witness that detects $\rho$. Our algorithm
is the best-known algorithm for the general separability problem;
it gets its advantage over the `basic algorithm' from reducing the
problem to an optimization over the pure separable states. If
properly implemented, the algorithm should give a feasible test
for separability in low dimensions (e.g. $MN<10$, with current
technology). The general technique depends only on the convexity
of the set of separable states, and thus can, in principle, be
applied to test for multi-partite entanglement. The algorithm also
gives experimentalists a tool for potentially determining if an
unknown state is entangled by measuring only a subset of the
expected values which completely describe the state. This method
effectively trades quantum resources (additional copies of $\rho$)
for classical resources (a computer able to calculate
$\mathcal{O}$).


We would like to thank Carolina Moura Alves, Coralia Cartis, and
Tom Stace for useful discussions.  We acknowledge support from the
EC under project RESQ (IST-2001-37559).  LMI, BCT, and DC also
acknowledge support from, respectively, CESG and NSERC; CMI; and
NSERC and the University of Waterloo.


\begin{thebibliography}{25}
\expandafter\ifx\csname natexlab\endcsname\relax\def\natexlab#1{#1}\fi
\expandafter\ifx\csname bibnamefont\endcsname\relax
  \def\bibnamefont#1{#1}\fi
\expandafter\ifx\csname bibfnamefont\endcsname\relax
  \def\bibfnamefont#1{#1}\fi
\expandafter\ifx\csname citenamefont\endcsname\relax
  \def\citenamefont#1{#1}\fi
\expandafter\ifx\csname url\endcsname\relax
  \def\url#1{\texttt{#1}}\fi
\expandafter\ifx\csname urlprefix\endcsname\relax\def\urlprefix{URL }\fi
\providecommand{\bibinfo}[2]{#2}
\providecommand{\eprint}[2][]{\url{#2}}

\bibitem[{\citenamefont{Nielsen and Chuang}(2000)}]{NC00}
\bibinfo{author}{\bibfnamefont{M.~A.}~\bibnamefont{Nielsen}} \bibnamefont{and}
  \bibinfo{author}{\bibfnamefont{I.~L.}~\bibnamefont{Chuang}},
  \emph{\bibinfo{title}{Quantum Computation and Quantum Information}}
  (\bibinfo{publisher}{Cambridge University Press},
  \bibinfo{address}{Cambridge}, \bibinfo{year}{2000}).

\bibitem[{\citenamefont{Gurvits}(2003)}]{Gur03}
\bibinfo{author}{\bibfnamefont{L.}~\bibnamefont{Gurvits}}, in
  \emph{\bibinfo{booktitle}{Proceedings of the thirty-fifth {ACM} symposium on
  Theory of computing}} (\bibinfo{publisher}{ACM Press}, \bibinfo{address}{New
  York}, \bibinfo{year}{2003}), pp. \bibinfo{pages}{10--19}.

\bibitem[{\citenamefont{Peres}(1996)}]{Per96}
\bibinfo{author}{\bibfnamefont{A.}~\bibnamefont{Peres}},
  \bibinfo{journal}{Phys. Rev. Lett.} \textbf{\bibinfo{volume}{77}},
  \bibinfo{pages}{1413} (\bibinfo{year}{1996}).

\bibitem[{\citenamefont{Horodecki and Horodecki}(1999)}]{HH99}
\bibinfo{author}{\bibfnamefont{M.}~\bibnamefont{Horodecki}} \bibnamefont{and}
  \bibinfo{author}{\bibfnamefont{P.}~\bibnamefont{Horodecki}},
  \bibinfo{journal}{Phys. Rev. A} \textbf{\bibinfo{volume}{59}},
  \bibinfo{pages}{4206} (\bibinfo{year}{1999}).

\bibitem[{\citenamefont{Nielsen and Kempe}(2001)}]{NK00}
\bibinfo{author}{\bibfnamefont{M.~A.}~\bibnamefont{Nielsen}} \bibnamefont{and}
  \bibinfo{author}{\bibfnamefont{J.}~\bibnamefont{Kempe}},
  \bibinfo{journal}{Phys. Rev. Lett.} \textbf{\bibinfo{volume}{86}},
  \bibinfo{pages}{5184} (\bibinfo{year}{2001}).

\bibitem[{\citenamefont{Doherty et~al.}(2002)\citenamefont{Doherty, Parrilo,
  and Spedalieri}}]{DPS01}
\bibinfo{author}{\bibfnamefont{A.~C.} \bibnamefont{Doherty}},
  \bibinfo{author}{\bibfnamefont{P.~A.} \bibnamefont{Parrilo}},
  \bibnamefont{and} \bibinfo{author}{\bibfnamefont{F.~M.}
  \bibnamefont{Spedalieri}}, \bibinfo{journal}{Phys. Rev. Lett.}
  \textbf{\bibinfo{volume}{88}}, \bibinfo{pages}{187904}
  (\bibinfo{year}{2002}).

\bibitem[{\citenamefont{Doherty et~al.}(2003)\citenamefont{Doherty, Parrilo,
  and Spedalieri}}]{qphDPS03}
\bibinfo{author}{\bibfnamefont{A.~C.} \bibnamefont{Doherty}},
  \bibinfo{author}{\bibfnamefont{P.~A.} \bibnamefont{Parrilo}},
  \bibnamefont{and} \bibinfo{author}{\bibfnamefont{F.~M.}
  \bibnamefont{Spedalieri}}, \emph{\bibinfo{title}{A complete family of
  separability criteria}} (\bibinfo{year}{2003}),
  \bibinfo{note}{quant-ph/0308032}.

\bibitem[{\citenamefont{Braunstein et~al.}(1999)\citenamefont{Braunstein,
  Caves, Jozsa, Linden, Popescu, and Schack}}]{BCJLPS99}
\bibinfo{author}{\bibfnamefont{S.~L.} \bibnamefont{Braunstein}},
  \bibinfo{author}{\bibfnamefont{C.~M.} \bibnamefont{Caves}},
  \bibinfo{author}{\bibfnamefont{R.}~\bibnamefont{Jozsa}},
  \bibinfo{author}{\bibfnamefont{N.}~\bibnamefont{Linden}},
  \bibinfo{author}{\bibfnamefont{S.}~\bibnamefont{Popescu}}, \bibnamefont{and}
  \bibinfo{author}{\bibfnamefont{R.}~\bibnamefont{Schack}},
  \bibinfo{journal}{Phys. Rev. Lett.} \textbf{\bibinfo{volume}{83}},
  \bibinfo{pages}{1054} (\bibinfo{year}{1999}).

\bibitem[{\citenamefont{Zyczkowski et~al.}(1998)\citenamefont{Zyczkowski,
  Horodecki, Sanpera, and Lewenstein}}]{ZHSL98}
\bibinfo{author}{\bibfnamefont{K.}~\bibnamefont{Zyczkowski}},
  \bibinfo{author}{\bibfnamefont{P.}~\bibnamefont{Horodecki}},
  \bibinfo{author}{\bibfnamefont{A.}~\bibnamefont{Sanpera}}, \bibnamefont{and}
  \bibinfo{author}{\bibfnamefont{M.}~\bibnamefont{Lewenstein}},
  \bibinfo{journal}{Phys.Rev. A} \textbf{\bibinfo{volume}{58}},
  \bibinfo{pages}{883} (\bibinfo{year}{1998}).

\bibitem[{\citenamefont{Gurvits and Barnum}(2002)}]{GB02}
\bibinfo{author}{\bibfnamefont{L.}~\bibnamefont{Gurvits}} \bibnamefont{and}
  \bibinfo{author}{\bibfnamefont{H.}~\bibnamefont{Barnum}},
  \bibinfo{journal}{Phys. Rev. A} \textbf{\bibinfo{volume}{66}},
  \bibinfo{pages}{062311} (\bibinfo{year}{2002}).

\bibitem[{\citenamefont{Bell}(1964)}]{Bel64}
\bibinfo{author}{\bibfnamefont{J.~S.} \bibnamefont{Bell}},
  \bibinfo{journal}{Physics} \textbf{\bibinfo{volume}{1}}, \bibinfo{pages}{195}
  (\bibinfo{year}{1964}).

\bibitem[{\citenamefont{Clauser et~al.}(1969)\citenamefont{Clauser, Horne,
  Shimony, and Holt}}]{CHSH69}
\bibinfo{author}{\bibfnamefont{J.~F.} \bibnamefont{Clauser}},
  \bibinfo{author}{\bibfnamefont{M.~A.} \bibnamefont{Horne}},
  \bibinfo{author}{\bibfnamefont{A.}~\bibnamefont{Shimony}}, \bibnamefont{and}
  \bibinfo{author}{\bibfnamefont{R.~A.} \bibnamefont{Holt}},
  \bibinfo{journal}{Phys. Rev. Lett.} \textbf{\bibinfo{volume}{23}},
  \bibinfo{pages}{881} (\bibinfo{year}{1969}).

\bibitem[{\citenamefont{G{\"{u}}hne et~al.}(2002)\citenamefont{G{\"{u}}hne,
  Hyllus, Bru\ss, Ekert, Lewenstein, Macchiavello, and Sanpera}}]{GHBELMS02}
\bibinfo{author}{\bibfnamefont{O.}~\bibnamefont{G{\"{u}}hne}},
  \bibinfo{author}{\bibfnamefont{P.}~\bibnamefont{Hyllus}},
  \bibinfo{author}{\bibfnamefont{D.}~\bibnamefont{Bru\ss}},
  \bibinfo{author}{\bibfnamefont{A.}~\bibnamefont{Ekert}},
  \bibinfo{author}{\bibfnamefont{M.}~\bibnamefont{Lewenstein}},
  \bibinfo{author}{\bibfnamefont{C.}~\bibnamefont{Macchiavello}},
  \bibnamefont{and} \bibinfo{author}{\bibfnamefont{A.}~\bibnamefont{Sanpera}},
  \bibinfo{journal}{Phys. Rev. A} \textbf{\bibinfo{volume}{66}},
  \bibinfo{pages}{062305} (\bibinfo{year}{2002}).

\bibitem[{\citenamefont{Barbieri et~al.}(2003)\citenamefont{Barbieri, Martini,
  Nepi, Mataloni, D'Ariano, and Macchiavello}}]{qphBMNMDM03}
\bibinfo{author}{\bibfnamefont{M.}~\bibnamefont{Barbieri}},
  \bibinfo{author}{\bibfnamefont{F.~D.} \bibnamefont{Martini}},
  \bibinfo{author}{\bibfnamefont{G.~D.} \bibnamefont{Nepi}},
  \bibinfo{author}{\bibfnamefont{P.}~\bibnamefont{Mataloni}},
  \bibinfo{author}{\bibfnamefont{G.~M.} \bibnamefont{D'Ariano}},
  \bibnamefont{and}
  \bibinfo{author}{\bibfnamefont{C.}~\bibnamefont{Macchiavello}},
  \emph{\bibinfo{title}{Experimental detection of entanglement with polarized
  photons}} (\bibinfo{year}{2003}), \bibinfo{note}{quant-ph/0307003}.

\bibitem[{\citenamefont{Horodecki et~al.}(1996)\citenamefont{Horodecki,
  Horodecki, and Horodecki}}]{HHH96}
\bibinfo{author}{\bibfnamefont{M.}~\bibnamefont{Horodecki}},
  \bibinfo{author}{\bibfnamefont{P.}~\bibnamefont{Horodecki}},
  \bibnamefont{and}
  \bibinfo{author}{\bibfnamefont{R.}~\bibnamefont{Horodecki}},
  \bibinfo{journal}{Phys. Lett. A} \textbf{\bibinfo{volume}{223}},
  \bibinfo{pages}{1} (\bibinfo{year}{1996}).

\bibitem[{\citenamefont{Horodecki and Ekert}(2001)}]{qphHE01}
\bibinfo{author}{\bibfnamefont{P.}~\bibnamefont{Horodecki}} \bibnamefont{and}
  \bibinfo{author}{\bibfnamefont{A.}~\bibnamefont{Ekert}},
  \emph{\bibinfo{title}{Direct detection of quantum entanglement}}
  (\bibinfo{year}{2001}), \bibinfo{note}{quant-ph/0111064}.

\bibitem[{\citenamefont{Carteret}(2003)}]{qphCar03}
\bibinfo{author}{\bibfnamefont{H.}~\bibnamefont{Carteret}},
  \emph{\bibinfo{title}{Noiseless circuits for the {P}eres criterion}}
  (\bibinfo{year}{2003}), \bibinfo{note}{quant-ph/0309216}.

\bibitem[{\citenamefont{Horodecki}(1997)}]{H97}
\bibinfo{author}{\bibfnamefont{P.}~\bibnamefont{Horodecki}},
  \bibinfo{journal}{Phys. Lett. A} \textbf{\bibinfo{volume}{232}},
  \bibinfo{pages}{333} (\bibinfo{year}{1997}).

\bibitem[{\citenamefont{Terhal}(2000)}]{Ter00}
\bibinfo{author}{\bibfnamefont{B.~M.} \bibnamefont{Terhal}},
  \bibinfo{journal}{Phys. Lett. A} \textbf{\bibinfo{volume}{271}},
  \bibinfo{pages}{319} (\bibinfo{year}{2000}).

\bibitem[{\citenamefont{Atkinson and Vaidya}(1995)}]{AV95}
\bibinfo{author}{\bibfnamefont{D.~S.} \bibnamefont{Atkinson}} \bibnamefont{and}
  \bibinfo{author}{\bibfnamefont{P.~M.} \bibnamefont{Vaidya}},
  \bibinfo{journal}{Mathematical Programming} \textbf{\bibinfo{volume}{69}},
  \bibinfo{pages}{1} (\bibinfo{year}{1995}).

\bibitem[{\citenamefont{Nesterov and Nemirovskii}(1994)}]{NN94}
\bibinfo{author}{\bibfnamefont{Y.}~\bibnamefont{Nesterov}} \bibnamefont{and}
  \bibinfo{author}{\bibfnamefont{A.}~\bibnamefont{Nemirovskii}},
  \emph{\bibinfo{title}{Interior-Point Polynomial Algorithms in Convex
  Programming}} (\bibinfo{publisher}{SIAM}, \bibinfo{address}{Philadelphia},
  \bibinfo{year}{1994}).

\bibitem[{\citenamefont{Renegar}(2001)}]{Ren01}
\bibinfo{author}{\bibfnamefont{J.}~\bibnamefont{Renegar}},
  \emph{\bibinfo{title}{A Mathematical View of Interior-Point Methods in Convex
  Optimization}} (\bibinfo{publisher}{MPS-SIAM},
  \bibinfo{address}{Philadelphia}, \bibinfo{year}{2001}).

\bibitem[{\citenamefont{Horst and Pardalos}(1995)}]{HP95}
\bibinfo{editor}{\bibfnamefont{R.}~\bibnamefont{Horst}} \bibnamefont{and}
  \bibinfo{editor}{\bibfnamefont{P.}~\bibnamefont{Pardalos}}, eds.,
  \emph{\bibinfo{title}{Handbook of Global Optimization}}
  (\bibinfo{publisher}{Kluwer Academic Publishers},
  \bibinfo{address}{Dordrecht}, \bibinfo{year}{1995}).

\bibitem[{\citenamefont{Hansen and Walster}(2004)}]{HW04}
\bibinfo{author}{\bibfnamefont{E.}~\bibnamefont{Hansen}} \bibnamefont{and}
  \bibinfo{author}{\bibfnamefont{G.}~\bibnamefont{Walster}},
  \emph{\bibinfo{title}{Global Optimization Using Interval Analysis}}
  (\bibinfo{publisher}{Marcel Dekker Incorporated}, \bibinfo{address}{Boston},
  \bibinfo{year}{2004}), ISBN \bibinfo{isbn}{0824740599}.

\bibitem[{\citenamefont{Curty et~al.}(2003)\citenamefont{Curty, Lewenstein, and
  L{\"{u}}tkenhaus}}]{qphCLL03}
\bibinfo{author}{\bibfnamefont{M.}~\bibnamefont{Curty}},
  \bibinfo{author}{\bibfnamefont{M.}~\bibnamefont{Lewenstein}},
  \bibnamefont{and}
  \bibinfo{author}{\bibfnamefont{N.}~\bibnamefont{L{\"{u}}tkenhaus}},
  \emph{\bibinfo{title}{Entanglement as precondition for secure quantum key
  distribution}} (\bibinfo{year}{2003}), \bibinfo{note}{quant-ph/0307151}.

\end{thebibliography}

\end{document}